\documentclass[two column,prl]{revtex4}
\usepackage{bm}
\usepackage{graphics}
\usepackage{graphicx}
\usepackage{amsfonts}
\usepackage{amsmath}
\usepackage{amssymb}
\usepackage{graphicx}

\begin{document}

\title[Short title for running header]{Hall helps Ohm: some corrections to negative-U centers approach to transport properties of YBa$_2$Cu$_3$O$_x$ and La$_{2-x}$Sr$_x$CuO$_4$}

\author{S.V. Baryshev}
\email{sergey.v.baryshev@gmail.com}
\affiliation{Argonne National
Laboratory, 9700 S. Cass Avenue, Argonne, IL 60439, USA}
\affiliation{Ioffe Institute, 26 Polytekhnicheskaya,
St.-Petersburg 194021, Russia}

\author{A.I. Kapustin}
\email{kapustinalex@yandex.ru} \affiliation{St. Petersburg State
Polytechnical University, 29 Polytekhnicheskaya, St. Petersburg,
195251, Russia}

\author{K.D. Tsendin}
\email{Tsendin@mail.ioffe.ru} \affiliation{Ioffe Institute, 26
Polytekhnicheskaya, St.-Petersburg 194021, Russia}
\affiliation{St. Petersburg State Polytechnical University, 29
Polytekhnicheskaya, St. Petersburg, 195251, Russia}


\begin{abstract}
For broad oxygen and strontium doping ranges, temperature
dependences ($T$-dependences) of the normal state resistivity
$\rho(T)$ of YBa$_{2}$Cu$_{3}$O$_{x}$ (YBCO) and
La$_{2-x}$Sr$_{x}$CuO$_{4}$ (LSCO) are calculated and compared to
experiments. Holes transport was taken in the
$\tau$-approximation, where $\tau$ is due to acoustic phonons.
Besides, $T$-dependence of the chemical potential $\mu(T)$ and
effective carrier mass $m^\ast$ $\sim$10-100 free electron masses,
obtained by negative-$U$ centers modelling the $T$-dependence of
the Hall coefficient, were used to calculate $\rho(T)$. In
addition, it is demonstrated that anisotropy of the cuprates does
not affect the calculated $T$-variation of neither Hall
coefficient nor $\rho$, but only rescale their magnitudes by
factors depending on combinations of $m_{ab}$ and $m_c$.
\end{abstract}

\maketitle

\section{Introduction}
\label{intro}
\par The negative-$U$ centers (NUC) conception is a way to
describe and understand properties of high-temperature cuprate
superconductors \cite{TsendinTc,Mitsen-Ivanenko}. One of the
approaches \cite{TsendinTc} on how to build such a center is based
on an idea proposed by P.W. Anderson \cite{Anderson} (Fig.1) and
the results of Kulik and Pedan \cite{Kulik}. It is described by
Hubbard Hamiltonian with a negative correlation energy
\cite{Kulik}. In terms of this approach, we have obtained
dome-shaped dependence of the critical temperature of the
superconducting transition, $T_c$, on the doping level
\cite{TsendinTc} and calculated nonmonotonic $T$-dependences of
the normal-phase Hall coefficient for YBCO and LSCO
\cite{TsendinHall}.

\section{Results, Discussion and Conclusion}
\label{sec:2}
\par Assuming that the electric field is directed
along the $z$ axis, in order to obtain the $T$-dependence of the
conductivity we solve the kinetic Boltzmann equation for holes in
the $\tau$-approximation. In the case of the hole parabolic
dispersion relation, the conductivity is as follows

\begin{equation}
\sigma(T)=\frac{\sqrt{2\cdot m^{*}}\cdot
e^{2}}{3\cdot\pi^{2}\cdot\hbar^{3}}
\int_{-\infty}^{0}\frac{1}{1+e^{\frac{\mu(T)-\varepsilon}{T}}}
\frac{\partial}{\partial\varepsilon}(\tau\cdot
(-\varepsilon)^{\frac{3}{2}})d\varepsilon.
\end{equation}

\begin{figure}[t] \centering
\includegraphics[width=5.5cm]{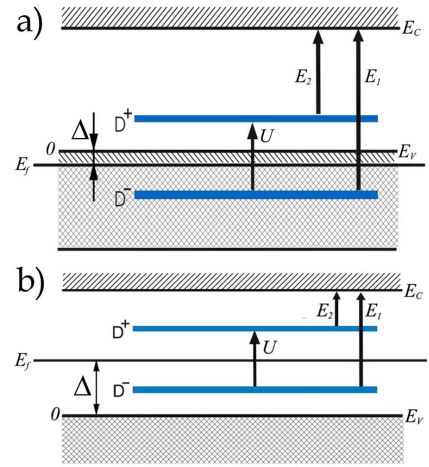}
\caption{Band diagram of cuprate HTSs in the NUC model: (a)
optimal doping mode with metallic conductivity ($\Delta
> 0$) and (b) low doping mode with semiconducting conductivity
($\Delta < 0$). $E_1$ and $E_2$ are the first and the second
ionization energies of the $D^-$ state, $U = (E_1 - E_2)$ is the
absolute value of the correlation energy, $D^+$ and $D^-$ are NUC
bands, and $\Delta = E_v - E_f$, where $E_v$ is the valence band
top and $E_f$ is the Fermi level.}\label{fig:1}
\end{figure}

\par Further, temperature dependence $\mu(T)$ along with effective
mass $m^\ast$ were extracted from numerical Hall concentration
calculations. In general, free model parameters were NUC
concentration $D$, $m^\ast$ and $\Delta$ (distance between the
valence band top and the Fermi level, see Fig.1)
\cite{TsendinHall,sust}. In accord with known $1/\tau\sim T$
\cite{Timusk}, scattering owing to acoustic phonons \cite{Anselm}
was considered

\begin{equation}
\tau(\varepsilon,T)=\frac{9\cdot\pi}{4\cdot\sqrt{2}}\cdot\frac{M\cdot
v_{0}^{2}\cdot \hbar^{4}}{\Omega_{0}\cdot C^{2}\cdot
(m^{*})^{3/2}\cdot k \cdot T}\cdot\frac{1}{\sqrt{\varepsilon}},
\end{equation}
where $M$ is the mass of a lattice ion, $C=\frac{\hbar^{2}}{2\cdot
m^{\ast}\cdot a}$ is the electron-phonon interaction constant, $a$
is lattice constant, $v_{0}$ is speed of sound of
$\sim$$10^{5}$--$10^{6}$ cm/s \cite{AcousticPhonons}, $\Omega_{0}$
is unit cell volume, $k$ is the Boltzmann constant, $\varepsilon$
is the energy.

\par Results of the resistivity temperature dependence calculations based on
Eqs. (1) and (2) taking into account $\mu(T)$ and $m^\ast$ are
demonstrated in Figs.2 and 3.

\begin{figure}[t] \centering
\includegraphics[width=5.5cm]{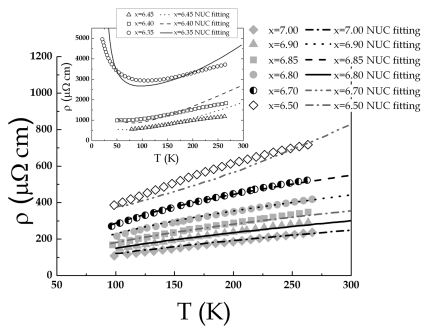}
\caption{$T$-dependence of the resistivity for the
YBa$_{2}$Cu$_{3}$O$_{x}$ system: symbols are experimental data
from Ref.\cite{JonesYBCO}, lines are fitting curves obtained in
terms of the NUC model.}\label{fig:1}
\end{figure}

\begin{figure}[t] \centering
\includegraphics[width=5.5cm]{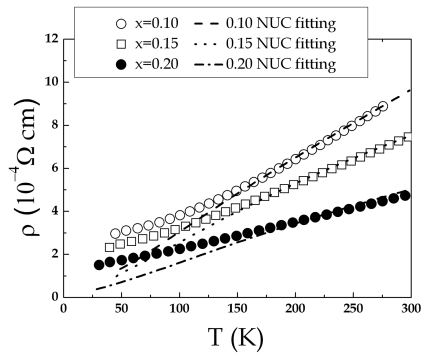}
\caption{$T$-dependence of the resistivity for the
La$_{2-x}$Sr$_{x}$CuO$_{4}$ system: symbols are experimental data
from Ref.\cite{SuzukiLSCO}, lines are fitting curves obtained in
terms of the NUC model.}\label{fig:1}
\end{figure}

\par In our previous studies, we assumed that the Fermi level moves
into the band gap at a certain doping level $x$. The Fig.3
demonstrates that $\rho(T)$ starts to show the semiconducting
behavior at increasingly high temperatures as the Fermi level
moves into the band gap ($\Delta$ becomes increasingly negative).
At $x$ = 6.35, this "switching" temperature is $\sim$110 K or
$\sim$9 meV (see also Ref.\cite{sust}).

\par In addition, generalizing the parabolic dispersion relation $\varepsilon(k)$ used for
the Hall and hence for the resistivity calculations to the
anisotropic case one can get only corrections in the form of
coefficients $(\frac{m_1}{m_2})^2$, where
$\frac{1}{m_1}=\frac{1}{3}\cdot(\frac{1}{m_c}+\frac{2}{m_{ab}})$
and $\frac{1}{m_2}=\sqrt{\frac{1}{3}\cdot(\frac{2}{m_c\cdot
m_{ab}}+\frac{1}{m_{ab}^2})}$, for the Hall coefficient and
$\sqrt[3]{\frac{m_c}{m_{ab}}}$ for $\rho$ (see also
Ref.\cite{Anselm}). Thus, anisotropy of charge carriers dispersion
does not introduce change in the $T$-run of the resistivity (and
$T$-dependence of the Hall coefficient as well). For both LSCO and
YBCO, $m^\ast$ was $\sim$10-100 free electron masses. This mass is
mostly of the same anisotropic origin being overestimated,
equalled $\sqrt[3]{m_{ab}^2\cdot m_c}$.

\par The comparison of the calculations results against experimental
data suggests that the NUC model consistently describes
$T$-dependences of the Hall coefficient and the resistivity of
cuprate HTSCs in the normal phase plus to correctly calculated
$T_c$ values.

\section{Acknowledgements}
\label{sec:3}
The study was funded by the RF Ministry of Education
and Science (project 2.1.1/988) and in part under contract
DE-AC02-06CH11357 between UChicago Argonne, LLC and the US DOE.

\end{document}